\documentstyle[prb,aps,twocolumn]{revtex}

\begin{document}
\draft
\title{About one forgotten mechanism of nonlinearity in the theory of hot
electrons}

\author{Yu. G. Gurevich, I. N. Volovichev\cite{AAAuth}}
\address{Departamento de F\'{\i}sica, CINVESTAV-I.P.N., Apdo. Postal
14-740, M\'exico D.F. 07000}
\date{\today}
\maketitle

\begin{abstract}
It is shown that in general electron gas  heating inevitably
results in the change of the carrier concentration in the
conduction band. It is proved, that this change, as a rule, leads
to the kinetic coefficient nonlinearity of the same order, as the
change of mobility does. The conditions are determined, when this
change can be neglected.
\end{abstract}

It is well known, that the fundamental reason for nonlinearity of
a current-voltage characteristic (CVC) of a homogeneous
semiconductor in strong electrical fields is the change of the
mean carrier energy (carrier heating). The classical theory of the
hot carrier transport was developed a long time ago and rather
explicitly \cite{Conwel,Book1,Grib,JR,Book2}. Thus, as a rule, it
was considered, that the nonlinearity of the CVC is related to the
carrier mobility alteration because of the change of carrier mean
energy. Some number of works was devoted to nonlinearity caused by
impact ionization \cite{Keld,Zi}, carrier lifetime change
\cite{AsheRec}, inter-valley redistribution of carriers \cite{Jorg}
or by non-parabolic form of the carrier dispersion law
\cite{Kane,Pogreb} in strong fields. In single-valley
semiconductors, neglecting such processes as an impact ionization
and carrier lifetime change in strong fields, it is usually
considered \cite{Conwel,Book1,Grib,JR,Book2} that only the
carriers, which already exist in bands, are subjected to heating,
i.e.\ during heating the concentration of carriers remains equal
to its value at the state of thermodynamic equilibrium. However,
as it was shown in Ref. \cite{Conwell1}, there exists one more
mechanism of nonlinearity, connected with the fact that the
violation of the energy equilibrium between electrons and holes
(the difference between electron and hole temperatures) inevitably
results in the violation of the concentration equilibrium between
electrons of the conduction band and the valence band. This idea
was developed in Ref. \cite{ConZuck}. However, due to the
assumption made that the population of the impurity level does not
depend on heating, the results turn out to be incorrect if the
heating of electrons and holes is different (electron and holes
temperatures are unequal). Besides, the question when this effect
takes place was left open.

In general, the discrepancy between electron and hole temperatures
should cause the change of the carrier concentrations in both
bands and the impurity level population. Hence, the problem is
reduced only to what is the magnitude of the contribution of the
latter effect to kinetic coefficients in comparison with the
change of mobility (relaxation time) at the same temperature
discrepancy. Thus, a new origin of strong field nonlinear effects,
related to the alteration of the energy level population in
conduction and valence bands due to the difference between
electron and hole temperatures in strong electrical fields, is
discussed below. It is possible to neglect this phenomena in the
theory of hot electrons only under special conditions indicated
below.

Experimental verification of the magnitude of the considered
effect is shown in Ref. \cite{Zucker1,Zucker2}. Unfortunately, in
spite of all these facts, the considered mechanism of nonlinearity
in the theory of hot electrons had been actually forgotten and did
not obtain a further development.

Virtually, the carrier concentration change because of this
mechanism is reduced to the carrier recombination rate change,
which is owing to the alteration of carrier distribution
functions.

Let's consider the elementary model --- a homogeneous
single-valley semiconductor with one nondegenerate impurity level
at an energy $\varepsilon_t$ and concentration of impurity $N_t$.
In the given model the kinetics of the carrier concentration
change within bands due to heating is determined by the following
processes: (1) the capture of electrons from conduction band to
the impurity level, (2) the thermal emission of electrons from the
impurity level into the conduction band, (3) the capture of holes
to the impurity level, (4) the thermal emission of holes from the
impurity level into the valence band. To simplify the calculations
we neglect the influence of interband transitions on carrier
concentration change at heating.

Let's note, that the concentration change is caused by the
alteration of the rate of recombination rather than thermal
generation.

Suppose, that only the electron subsystem is subjected to heating.
Then, if conditions for carrier concentrations indicated in
\cite{FP} are fulfilled, the electron gas can be described
by the Fermi distribution function with electron temperature
$T_e$. Subsystems of holes and captured carriers have the lattice
temperature $T_0$.

The capture rate of electrons onto the impurity level can be
represented by following expression \cite{Newm}:

\begin {equation}
r_n = \alpha_n(T_e)N_t[1-f_t(T_0)]n(T_e),
\end {equation}
where $f_t$ is the distribution function of electrons on the
impurity level, $N_t[1-f_t(T_0)]$ represents the concentration of
free impurity states, $n(T_e)$ is the concentration of electrons,
$\alpha_n(T_e)$ is the capture factor of electrons by the trap.
Let's emphasize once again, that here we do not consider the
explicit dependence of the capture factor on the magnitude of the
electrical field applied, i.e.\ we do not take into account such
processes as the change of the carrier life time in strong fields
(see \cite{AsheRec}). By definition,
\begin{equation}
\alpha_n(T_e) = \frac{\int^\infty_{\varepsilon_c}c_n(\varepsilon)
\nu_n(\varepsilon)f_n(\varepsilon,T_e)d\varepsilon}
{\int^\infty_{\varepsilon_c}\nu_n(\varepsilon)f_n(\varepsilon,T_e)
d\varepsilon}.
\end{equation}
Here $c_n(\varepsilon)$ is a probability of the electron
transition from the impurity level into the state with an energy
$\varepsilon$, $\nu_n(\varepsilon)$ is a density of states,
$f_n(\varepsilon,T_e)$ is the Fermi distribution function with
temperature $T_e$ for conductivity electrons, $\varepsilon_c$ is
the energy of an electron at the bottom of conduction band.

The rate of a thermal emission into the conduction band is assumed
to be independent of the temperature of electrons in the
conduction band (that is correct, at least, for wide-band
semiconductors):
\begin{equation}
g_{nT} = \alpha_n(T_0)N_tf_t(T_0)n_1,
\end{equation}
where $n_1\equiv\nu_{n0}\exp\left(-{\cal I}/T_0\right)$ is a
parameter describing the impurity level, $\nu_{n0}$ is the
effective density of states in the conduction band, ${\cal
I}\equiv\varepsilon_c-\varepsilon_t$ is the ionization energy of
the impurity level. The parameter $n_1$ represents concentration
of electrons in the conduction band, which would take place, if
the Fermi level would coincide with the impurity level.

Obviously, the recombination rate of electrons is equal to
$R_n=r_n-g_{nT}$.

Similar equations can be written for the hole subsystem as well:
\begin{eqnarray}
&&r_p=\alpha_p(T_0)N_tf_t(T_0)p,\\
&&g_{pT}=\alpha_p(T_0)N_t[1-f_t(T_0)]p_1,
\qquad R_p=r_p-g_{pT}.
\end{eqnarray}
Here $p$ is a hole concentration, the definition of quantities
$\alpha_p(T_0)$ and $p_1$ is similar to the above mentioned one
for electrons.

In a homogeneous semiconductor in steady state $R_n=R_p=0$, and
the condition of the electroneutrality is fulfilled. It is
convenient to express such a condition as:
\begin{equation}
\label{neutral}
\delta n + N_t\delta\! f_t = \delta p,
\end {equation}
where $\delta n\equiv n-n_0$, $\delta p\equiv p-p_0$, $\delta\!
f_t\equiv f_t-f^0_t$, and $n_0$, $p_0$ and $f^0_t$ are
respectively electron and hole concentrations and the impurity
level population in the absence of heating ($T_e=T_0$).

After solving the set of equations $R_n=R_p=0$ together with the
Eq. \ (\ref{neutral}), one obtains the following expressions for
the carrier concentration change caused by heating:
\begin{eqnarray}
\label{dn0}
\nonumber
\delta
n&=&-\left[1+\frac{p_1(n_0+n_1)^2}{n_1(p_0+p_1)^2+N_tn_1p_1}\right]^{-1}\\
&&\times\frac{n_0}{\alpha_n(T_0)}\frac{\partial
\alpha_n(T_0)}{\partial T}\delta T,\\
\label{dp0}
\delta p&=&\delta n\left[ 1+\frac{N_tp_1}{(p_0+p_1)^2}\right]^{-1},
\qquad \delta T\equiv T_e-T_0\ll T_0.
\end{eqnarray}

Assuming additionally, that the gas of carriers is nondegenerate,
i.e.\ the relation $n_0p_0=n_1p_1=n_i^2$ holds, where $n_i$ is the
intrinsic carrier concentration  (i.e.\ at $N_t=0$) in the absence
of heating, and, besides, bands are parabolic, we will analyze
Eqs.\ (\ref{dn0})--(\ref{dp0}) in two limiting cases.

1. An intrinsic semiconductor with low concentration of shallow
traps for electrons .\\ In this case $n_1\gg n_0=p_0=n_i\gg p_1$
and Eqs.\ (\ref{dn0})--(\ref{dp0}) acquire the form:
\begin{eqnarray}
\label{dni}
\frac{\delta n}{n_i}&=& - \frac{\partial \alpha_n(T_0)}{\partial T}
\frac{\delta T}{2\alpha_n(T_0)},\\
\label{dpi}
\delta p&\approx &\delta n.
\end{eqnarray}

Thus the magnitude of the carrier concentration change at heating
in this case is determined only by the temperature dependence of
the electron capture factor $\alpha_n$.

Let's note also, that expressions (\ref{dni})--(\ref{dpi}) have
the same form as in the case of only interband transition in an
intrinsic semiconductor.

2. A $n$-type monopolar semiconductor with donor impurity ($n_0\gg
n_i\gg p_0$).\\
The relative carrier concentration change in this
case is described by the following expressions:
\begin{eqnarray}
\label{dnn}
\frac{\delta n}{n_0}&=& -\left[1+1/\left({n_i^2\over
n_0^2}+{N_t\over n_1}\right)\right]^{-1} \frac{\partial
\alpha_n(T_0)}{\partial T}\frac{\delta
T}{\alpha_n(T_0)},\\
\label{dpn}
\delta p&=&\delta n\left[
1+\frac{n_0^2}{n_i^2}\frac{N_t}{n_1} \right]^{-1}.
\end{eqnarray}

It is easy to verify, that in this case the carrier concentration
change at heating is determined not only by the temperature
dependence of the electron capture factor $\alpha_n$, but also
depends on concentration of impurity and on the temperature $T_0$.
In the range of high temperatures ($T_0\gg {\cal
I}/\ln(\nu_{n0}/N_t)$, i.e.\ $N_t\ll n_1$) the functional
dependence of $|\delta n/n _ 0|$ on $N_t$ has a deep minimum at
$N_t\approx\sqrt[3]{n_i^2n_1}$, being equal to
\begin{eqnarray}
\nonumber
\left|{\delta n\over n_0}\right|_{\rm min} &=& \left({2n_i\over
n_1}\right)^{2/3}\frac{\partial \alpha_n(T_0)}{\partial
T}\frac{\delta T}{\alpha_n(T_0)}\\
&&\sim\exp\left(-{\varepsilon_g-2{\cal I}\over 3T}\right)\ll 1,
\end{eqnarray}
where $\varepsilon_g $ is the bandgap width. Thus, near the
indicated concentration the additional contribution to the
conductivity change is negligibly small and heating effects are
described by existing theories \cite{Conwel,Book1,Grib,JR,Book2}.

At low temperatures or, that is equivalent, heavily doping
($N_t\gg n_1$) we come back to expression (\ref{dni}). Let's note,
that in this case (in contrast to Eq.\ (\ref{dpi})) $\delta p\ll
\delta n$.

In monopolar case, as well as for an intrinsic semiconductor, the
deviation from the conventional theories is determined by the
temperature dependence of the electron capture factor $\alpha_n$.
For simplicity we will analyze model dependence for two cases:
attracting and repulsive potentials of impurity centers.

In the case of the electron capture by an attracting potential
$\alpha_n(T)\sim T^{-m}$, and $m$ varies within the limits from
$m\simeq 1$ up to $m\simeq 5$ depending on the nature of the
semiconductors and impurities
\cite{APY,Atr}. Then
\begin{equation}
\frac{\delta T}{\alpha_n(T_0)}\frac{\partial
\alpha_n(T_0)}{\partial T}=-m\frac{\delta T}{T_0}.
\end{equation}

In the case of electron capture by a repulsive potential of
impurity (for example, ions of gold or copper in germanium) the
temperature dependence of the capture factor is satisfactorily
described by the expression \cite{Bbonch}:
\begin{equation}
\alpha_n(T)\sim \exp\left[ -\left( {T^*\over
T}\right)^{1/3}\right],
\end{equation}
where $T^*$ is a parameter depending on the specific kind of the
semiconductor and the impurity in it.

Hence,
\begin{equation}
\frac{\delta T}{\alpha_n(T_0)}\frac{\partial
\alpha_n(T_0)}{\partial T}=\frac{1}{3} \left( {T^*\over
T_0}\right)^{1/3}\frac{\delta T}{T_0}.
\end{equation}

As long as $T^*$ usually lies in the range from $\rm 10^4\ K$ to
$\rm 10^9\ K$ \cite{Bbonch,Rep}, in the case of a repulsive
potential the contribution to the conductivity change in the
heating field can be even more essential.

Thus, the electron concentration change caused by carrier heating
alters kinetic coefficients as much as the mobility (relaxation
time) change does. It should be mentioned that there exist only
two specific, not very interesting, situation, namely: (1) the
electron Fermi level lies far enough both from the middle of the
gap and from the impurity level, and (2) the electron capture
factor weakly depends on the temperature, when the traditional
theory of hot electrons is correct.

We wish to thank Dr. F. P\'erez-Rodr\'{\i}guez for helpful
discussion. This work has been partially supported by
CONACyT--M\'{e}xico.

\end{document}